# Mechanism and modulation of terahertz generation from a semimetal - graphite

**Tong Ye[1], Sheng Meng[1], Jin Zhang[1], Yiwen E[1], Yuping Yang[2], Wuming Liu[1], Yan Yin[1,3,*], Li Wang[1,3]**

*1, Beijing National Laboratory for Condensed Matter Physics, Institute of Physics, Chinese Academy of Sciences, Beijing 100190*

*2, School of Science, Minzu University of China, Beijing, 100081*

*3, Cooperative Innovation Centre of Terahertz Science, Chengdu, Sichuan, 610054*

*\*Email address: yan.yin@aphy.iphy.ac.cn*

**Abstract:**

Semi-metals might offer a stronger interaction and a better confinement for terahertz wave than semiconductors, while preserve tunability. Particularly, graphene-based materials are envisioned as terahertz modulators, filters and ultra-broadband sources. However, the understanding of terahertz generation from those materials is still not clear, thus limits us recognizing the potential and improving device performances. Graphite, the mother material of graphene and a typical bulk semi-metal, is a good system to study semi-metals and graphene-based materials. Here we experimentally modulate and maximize the terahertz signal from graphite surface, thus reveal the mechanism - surface field driving photon induced carriers into transient current to radiate terahertz wave. We also discuss the differences between graphite and semiconductors; particularly graphite shows no temperature dependency from room temperature to 80°C. Above knowledge will help us understand terahertz generations, achieve maximum output and electric modulation, in semi-metal or graphene based devices.

As a fairly new frequency regime, terahertz (THz), and its related sciences and techniques have been heavily studied.[1-16] The generations of THz have been reported from and studied for semiconductors[17-20], superconductor[21], optical crystals[22,23], and plasma[24,25]. Semiconductors are most studied and used material for THz generations. The mechanisms for semiconductors are mainly considered to be well known and attributed to surface field effect[17,18], photo-Dember effect[19,20] or optical rectification. Metals are normally considered to be perfect mirrors at THz regime, and have no THz generation. Semi-metals might offer a stronger interaction with THz wave than semiconductors, while maintaining certain tunability. So far, the interactions between THz waves and semi-metallic materials are not well understood, including the THz generation from semi-metals. Particularly, researchers have proposed to use graphene or bi-layer graphene to make tunable terahertz metamaterials,[26] optical modulators,[27] terahertz tunable filters,[28] and ultra-broadband terahertz sources.[29] Because of the thin size and high carrier mobility of graphene, they might be candidates for nano-scale and ultra-broadband terahertz sources. However, reports of terahertz generations from graphene give different mechanism explanations, some attributing to photon drag effect[30,31] while another concluding to photo-Dember effect.[32] This contradiction highlights our shortfall in understanding the physics of terahertz generation from graphene related materials; at same time, this lack of knowledge limits us recognizing the potential or improving device performances, for instance maximizing the output or achieving electric modulation.

Graphite, mother material of graphene and a typical bulk semi-metal, can be a show case system to study the THz generation from graphene related materials or semi-metals in general. In some sense, understanding the process in graphite is equally or more important than that in graphene, because the generation mechanism in graphite is orders of magnitude stronger than those in graphene. When the number of stacking layers of graphene gets to a couple of tens, we believe it will behave like graphite in terms of THz generation. The terahertz generation of graphite was first observed by G. Ramakrishnan et al. in 2009, and was attributed to the transient photocurrent in the direction normal to the basal plane.[33] M. Irfan et al. reported that the THz signals from two different kinds of graphite samples with opposite doping had opposite phase.[34] A phenomenological description, (second or third order) nonlinear optical rectification,[33,35] is used to explain experimental results. However, how the carriers become transient current and the details of the mechanism are still unanswered.

Here, we report on the modulation and maximization of terahertz pulses radiated from voltage biased graphite. Our result clearly proves the physical nature of THz generation in graphite, surface field driving photon induced carriers into transient current to radiate THz wave. We also perform ab initio simulations to verify the gating field effect on graphite. The calculation result supports the gating modulation and saturation behavior, observed from our experiment. It also quantitatively supports that the THz saturation amplitude difference between positive and negative gate voltages is the result of the effective mass difference between electrons and holes, which have a ratio about 2.3. Compared to THz generation from semiconductors, the graphite shows differences: much quicker amplitude saturation for the surface field tuning, and no temperature dependence. All these two differences are the results of a much higher free-carrier density in graphite, a semi-metal, than semiconductors. This work not only helps us understand the physics of THz generations in bulk semi-metals or graphene based devices, but also demonstrates the electric modulation ability and signal maximization in semi-metallic THz generation devices.

**Results**

**Electrically modulated THz generation setup.** Figure 1a shows the sketch (side view) of the graphite sample and experiment setup. The graphite sample is a ZYA grade Highly Ordered Pyrolytic Graphite (HOPG) with a size of 12×12×2 mm$^3$ from Structure Probe Inc. Before THz emission measurements, a fresh flat surface was prepared by mechanical exfoliation. An ion-gel top gate was made for the application of a tunable electric field normal to and at the graphite surface. A Keithley 2400 is used as the gate voltage source ($V_g$) and monitors the gate current as well. An aluminum plate supports and connects the back side of the graphite to the negative electrode of the gate source. A circle of copper adhesive foil functions as the positive electrode of the gate voltage, is separated by a layer of insulting adhesive tape from the graphite, and is connected by ion-gel[36] with the top surface of the graphite. The resulting gate capacitance of the setup is about 1.3μF. The center openings of the copper adhesive foil and insulting tape are about 25 mm$^2$ in squares, in order for the placing of ion-gel on and the optical addressing to the graphite top surface.

A linear polarized femtosecond laser, operating at 800 nm central wavelength with 80 MHz repetition rate and 70 fs pulse duration, was focused onto the graphite surface. The incident laser power was about 400mW, the incident angle was 60°, and the spot diameter was about 2mm. The radiated THz pulses in the reflective direction were measured with a ZnTe (110) electro-optic crystal, and the detection 800nm laser power was about 30mW. The measured THz waveforms and their peak-to-peak amplitudes as functions of the gate voltage ($V_g$) are plotted in Fig. 2.

**Surface field modulation.** Fig. 2 clearly illustrates the modulation of the THz signals with the gate voltage. Fig. 2a shows the temporal waveforms of the THz signals generated from the basal plane of graphite, when the gate voltage decreases from positive 2.5V to negative 3V. Fig. 2b shows the peak-to-peak amplitude of the THz pulse as a function of the gate voltage. The gate voltage was scanned from 0V to 2.5V, and then was scanned back to -3V. Fig. 2b shows the data from both the voltage increasing scan (black square points) and decreasing scan (red circle points). In Fig. 2b, the gate voltage upward scan curve matches the downward scan curve very well; this indicates hysteresis of the gate scan is negligible. Therefore, we only plot the $V_g$ downward scan THz waveforms in Fig. 2a, which clearly shows that a higher gate voltage will give a stronger THz signal until saturation. Changing the gate voltage direction will result in the THz signal turning to the opposite direction. The bias voltage for minimum THz amplitude is not exactly 0V, because of a residue or chemical doping of the graphite surface. In Fig. 2b, we also notice that the THz pulses under different gating directions show different saturation amplitudes. We define $I_p$, the saturation amplitude for positive $V_g$, and $I_n$, the saturation amplitude for negative $V_g$. $I_p$ is 2.3 times of $I_n$, or, $I_p/I_n = 2.3$. This difference is due to the effective mess difference of electrons and holes in Z direction of graphite. We will prove this point later with our theoretical calculation.

Fig. 2, together with previously reported experiments,[33,37] clearly proves the mechanism of graphite THz generation. The excitation 800nm laser pulse is absorbed by the graphite surface to generate electron hole pairs; then, a surface field normal to the graphite basal plane drives the pairs into transient current to generate the THz radiation. A previous report shows that the THz signal is independent to the excitation light's polarization at various conditions (e.g. different incident angles),[37] and supports that the excitation pulse contributes the photo-carriers through sample light absorption. Here, our gating measurement proves that the surface field normal to the graphite basal plane is the driving force of the transient current, which radiates the THz pulse. Fig. 1b and 1c show the sketch of the surface field and electric potentials of graphite under different $Vg$, (b) for a large positive $Vg$, and (c) for a large negative $Vg$. When the $Vg$ is positive, it generates

an electric field pointing into and at the surface of graphite with the help of the ion-gel. This field penetrates into the graphite surface, thus builds a surface field normal to the basal plane. However, the field inside graphite is reduced due to the screening of graphite as it goes deep into the body. When the $Vg$ increases, the field on and inside graphite increase as well; the ion-gel will not fail in working until $Vg$ is outside the range of ±3V. When the $Vg$ is negative, the ion-gel works similarly but with the opposite field direction. When $Vg$ =0, graphite still has a small surface field because graphite is naturally n-doped. Therefore, we achieve the modulation of the surface field of graphite using ion-gel, and our result in Fig. 2 proves the direct connection between the surface field and the generated THz signal, both in magnitude and direction. Gopakumar et al. first attributed the terahertz generation of graphite to the transient photocurrent in the direction normal to the basal plane;[33] our work here, with well controlled experiment conditions, clearly demonstrates that how the photo-carriers are driven into transient current and the key role of surface field in the process.

**Simulation parameters and results.** We use ab initio theoretical method to calculate the surface field and electron dispersion curve of graphite in Z direction, prove that the saturation behavior of THz amplitude in Fig. 2b is the result of graphite screening of the external field, and quantitatively verify that the difference in saturation amplitudes for opposite field directions is indeed due to the effective mess difference of electrons and holes. The band structure and local potential of graphite are calculated based on density functional theory (DFT) with the Vienna Ab initio Simulation Package (VASP).[38] Projector-augmented wave pseudo-potential and plane wave energy cutoff of 600 eV are employed. Generalized gradient approximation in the PBE form[39] is used to express the exchange-correlation functional. The K-point grid of 15×15×5 is used to sample the first Brillouin zone of bulk graphite. The atomic structure is relaxed until the force on each atom is less than 0.01 eV/Å. The local potential in graphite in response to external electric field is calculated using a 10-layer graphite slab. When calculating a graphite slab, a sufficiently large vacuum space of more than 15 Å in the surface normal direction is adopted. An external field is applied using a dipole placed in the middle of vacuum region. The induced field strength inside graphite as a function of external field applied is calculated and shown in Fig. 3a. It is clear that the field at the surface layer of graphite (the outmost layer of the graphite slab) has approximately a linear relationship with the external field applied with a dimensionless slope of ~0.3. On the contrary, the field on layers below the first layer initially increases linearly with the external field $E_{ext}$ up to

$0.4 \times 10^{10}$ V/m, then it quickly saturates around a value of $0.02 \times 10^{10}$ V/m for $|E_{ext}| \geq 0.5 \times 10^{10}$ V/m. Under a positive field (with field vector outgoing from the graphite surface), although the induced field at the second layer keeps increasing, local field at layers below the second layer all slightly decreases with the increase in external field after $E_{ext} \geq 0.5 \times 10^{10}$ V/m. We also find that under the negative field (with field vector going into the graphite) the local fields at layers below the first layer have almost the same values, showing a stronger screening under negative field, probably due to more charges transferred to the first layer.

The calculated band structure along the K-H direction (Z direction, normal to basal plane) is shown in Fig. 3b. We find that besides the flat band around Fermi level, there are two additional dispersive electronic bands, which separate by ~1.4 eV at K point, and join each other at the Fermi level at H point. These two bands come from the interlayer interaction between the $p_z$ orbitals of graphite layers, which are sensitive to the interlayer distance of graphite. We find that the curvature of the two bands changes dramatically as the electron potential deviates from the Fermi level. The effective electron/hole masses are thus calculated from the curvatures of the two bands using the equation $m^* = \frac{\hbar^2}{\partial^2 E/\partial k^2}$, and are shown in the lower panel of Fig. 3b.

**Temperature dependence.** Furthermore, we look into the similarities and differences between graphite, a semi-metal, and semiconductors in the THz generation (surface field induced). Both of them generate THz with a surface field in the normal direction, and both of them show that a surface field tuning results in a THz signal modulation.[40] However, graphite THz generation appears with a much quicker amplitude saturation for the surface field tuning (in Fig. 2b), in comparison to the semiconductors.[40] Graphite also shows no temperature dependence, as semiconductors give highly enhanced THz radiation at a high temperature.[41] Fig. 4 shows the THz waveforms of the graphite surface at different sample temperatures. The graphite was heated by a small heater on the backside and the temperature was monitored with a sensor on the graphite front surface. Other measurement settings were the same as those of the gating measurement, but without the ion-gel on graphite for the $V_g$=0V measurement. The slight amplitude drops at high temperatures in Fig. 4 are due to an off-focus effect, as the mounting of the graphite sample extends about 0.3mm at the highest temperature. For the $V_g$=3V data, the THz signal drops intensity at 28°C, then goes to opposite phase at 46°C, and stays about same for further higher temperatures. This phase reversion for $V_g$=3V, probably is due to graphite's naturally n doping surface field overcomes and surpasses the ion-gel induced gating field. Either graphite's naturally n doping surface field gets enhanced or the ion-gel starts to fail, when the sample temperature goes higher from room temperature. In general, Fig. 4 demonstrates that the THz signal from graphite is not responding to the temperature change from room temperature to 80°C. All those two differences, the faster amplitude saturation and no temperature dependency, are the results of a much higher free-carrier density in graphite, a semi-metal, than semiconductors. The high free-carrier density provides a strong field screening in the surface field tuning measurement and makes graphite much less sensitive to temperature than semiconductors, whose defect associated carrier levels are normally very sensitive to the temperature.

**Discussion**

Fig. 3a indicates a strong screening effect exists in graphite: within 2-3 layers beneath the graphite surface, the external field is largely screened out and the local field saturates quickly around a relatively small value (2% of the external field). This field screening behavior in Fig. 3a is consistent with the saturation behavior of THz amplitude in Fig. 2b; this proves the saturation of the THz amplitude is the result of screening of the surface field. Considering the saturation field in graphite is about $0.02 \times 10^{10}$V/m and surface field extends only about 3 atom layers into the bulk (layer spacing is 0.335 nm), we have a maximum Fermi level of $\pm 0.2$eV at graphite surface while the bulk body is always considered to be the Fermi level zero point. When $V_g$ is a sufficiently large positive value, graphite surface (with ion-gel) has a potential of $V_{sur} \approx +0.2$eV (in Fig. 1b). This positive $V_{sur}$ corresponds to -0.2eV Fermi level in electron dispersion curve, and cuts the lower band (the blue one in Fig. 3b upper panel). The cutting position is the Fermi surface, determines the effective mass of the quasi-particles, and gives an effective mass m* = $2.5m_0$ ($m_0$ is the electron mass). For a large negative $V_g$ value (in Fig. 1c), $V_{sur} \approx -0.2$eV; the Fermi level cuts the upper band (the red one in Fig. 3b) at +0.2eV level, and gives an effective mass m* = $5.5m_0$. It is reasonable to assume that the THz amplitude is proportional to $E/m^*$, where E is the surface field and m* is the quasi-particle effective mass. Considering the saturation surface field is same for positive and negative $V_g$, we have $\frac{I_p}{I_n} = \frac{m^*|_{Vsur=-0.2eV}}{m^*|_{Vsur=+0.2eV}} = \frac{5.5}{2.5} = 2.2$. This theoretical value is consistent with our experimental value of 2.3 in Fig 2b. The agreement quantitatively proves that the difference in saturation amplitudes for opposite field directions is due to the effective mess difference of electrons and holes, and greatly supports our physical picture for the THz generation mechanism.

In conclusion, we carry out a surface field modulation and a temperature variation experiments on THz signal emitted from graphite. Our result clearly proves the THz generation mechanism, the surface field driving photo-carriers into transient current to radiate THz wave. We also perform ab initio calculation for the field screening effect and electron dispersion curves in Z direction of graphite. Theoretical result strongly supports our physical picture and shows that external field is screened after 2-3 atom layers of graphite. The THz saturation amplitudes of positive and negative gate voltages have a ratio of 2.3, because the effective mass of electrons is about 2.3 times of the effective mass of holes in Z direction (normal to basal plane) of highly gated graphite. Despite the similarities in the mechanism to semiconductors, graphite has quicker amplitude saturation and is insensitive to the temperature in the range of room temperature to 80°C.

**Acknowledgements**

This work was supported by the National Basic Research Program of China (No. 2014CB339800), by the National Natural Science Foundation of China (Grant No.11574388, 11574408), and by the Scientific Research Start-up Fund for the Returned Overseas Chinese Scholars from Ministry of Education of China.

**Author contributions**

T. Y. prepared and measured samples and devices; S. M. and J. Z. preformed the simulations, and wrote partial manuscript; Y. E prepared and operated the measurement system; Y. Yang and W. L. contributed in discussion; Y. Yin designed, coordinated the project and wrote the manuscript with the contributions from all other authors; L. W. provided the measurement system and contributed in discussion.

**Competing Financial Interests Statement** The authors declare no competing financial interests.

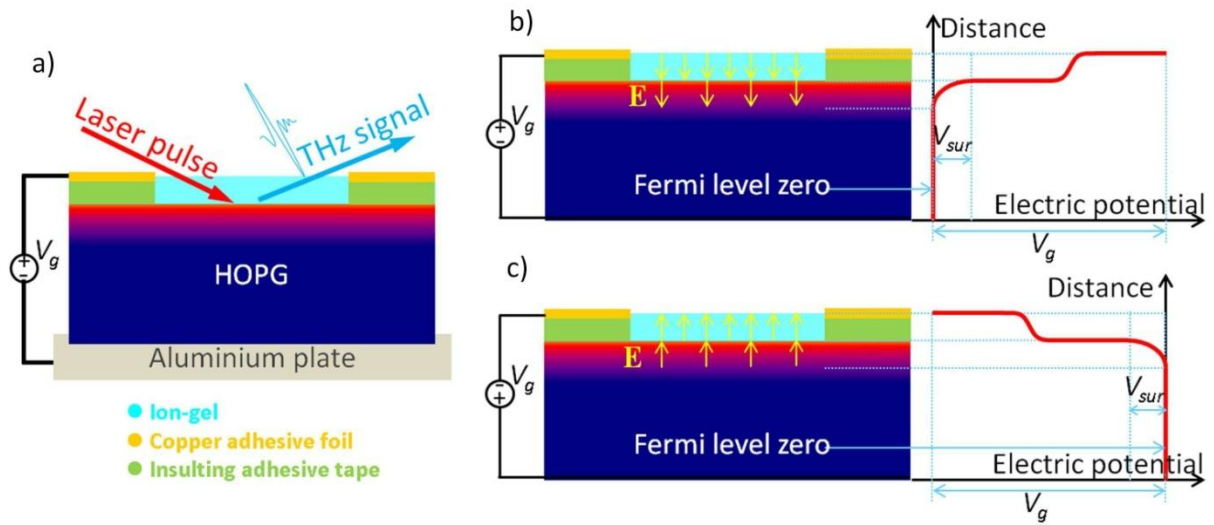

**Figure 1 | The sketch of sample, electrically modulated THz generation setup, the surface field and electric potentials of graphite under different $V_g$.** (a) The incident laser light (the red arrow) excites the top surface of the graphite, and the THz radiation (the blue arrow) is collected in the reflective direction. An ion-gel top gating method with gate voltage ($V_g$) is setup to modulate the surface field of the graphite. (b) The sketch of the surface field and electric potentials of graphite under different $V_g$ for a large positive $V_g$, and (c) for a large negative $V_g$.

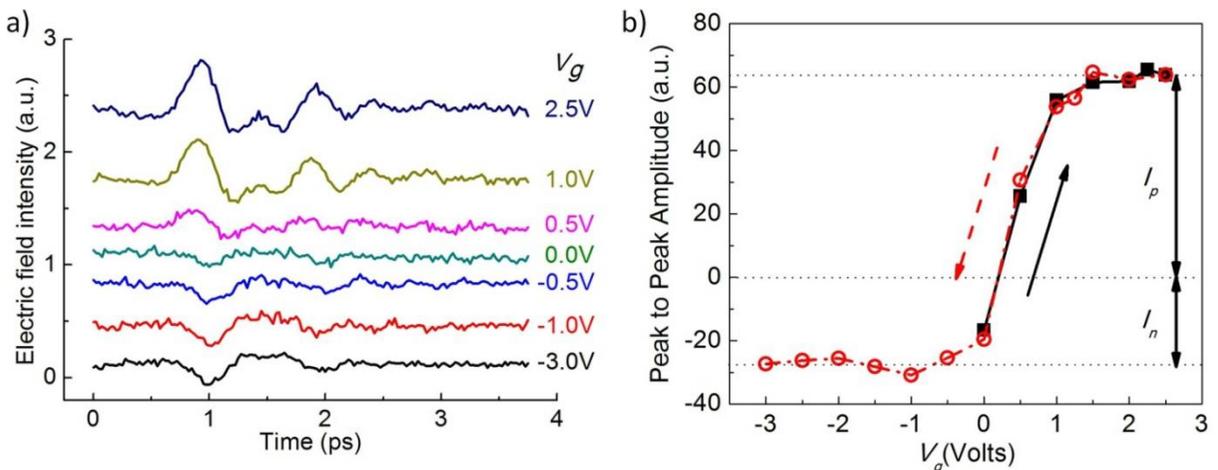

**Figure 2 | (a) THz waveforms and (b) peak-to-peak amplitudes as functions of the gate voltage ($V_g$).** In sub-figure (a), difference color curves represent the THz waveforms from graphite under different $V_g$ (corresponding values listed on the right side) during $V_g$ downward scan. In sub-figure (b), the black square data points are from $V_g$ upward scan; the red circle data points are from $V_g$ downward scan.

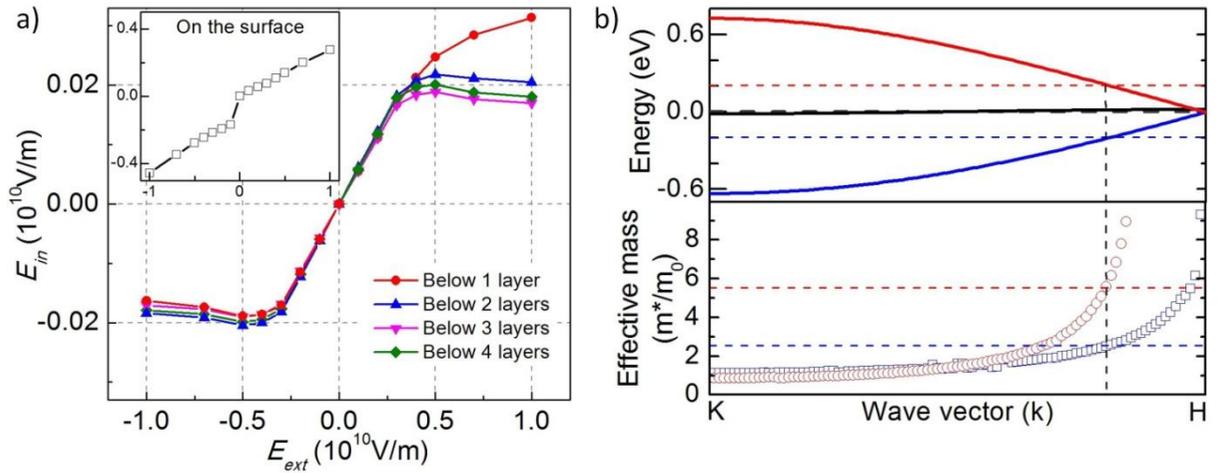

**Figure 3 | (a) Calculated electric field under graphite surface and on the surface (insert figure) as functions of the external electric field.** The calculation is from ab initio simulation for 10 graphite atom layers. **(b) Calculated electron dispersion curve (upper panel) and effective mass (lower panel) in Z direction (normal to basal plane) of graphite.**

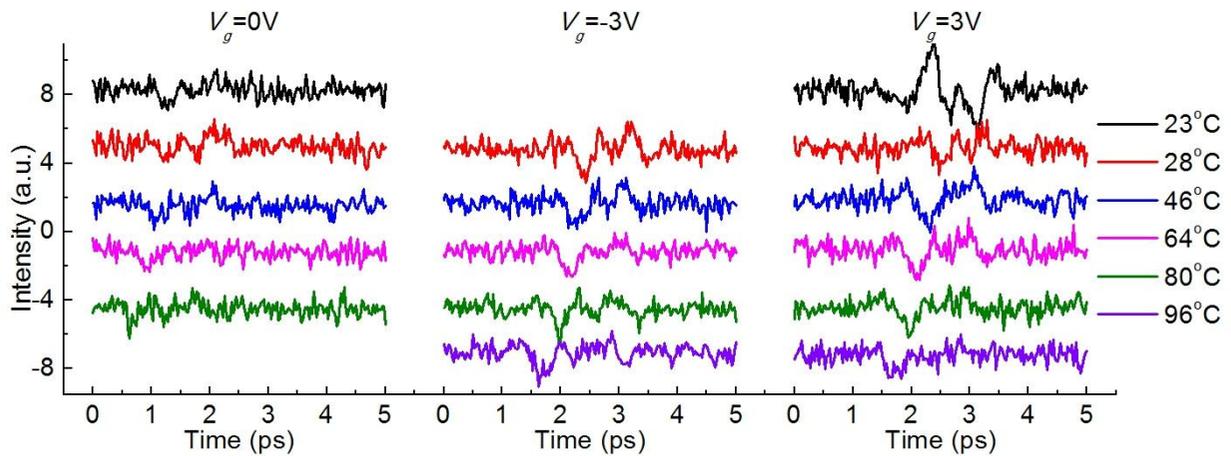

**Figure 4 | The THz waveforms as functions of the sample temperature.** From left to right, those curves are graphite surface emitted THz waveforms at different temperatures (plotted as different colors) for $V_g$=0V, $V_g$=-3V, $V_g$=3V, respectively. The colors of the curves represent the sample temperatures which are listed at the right side.